\documentclass[aps,twocolumn,showpacs,fleqn]{revtex4}

\usepackage{amsmath,amssymb}
\usepackage{graphicx}

\def\mrm#1{\mathrm{#1}}

\def\fig#1{Fig.\,\ref{#1}}
\def\figs#1{Figs.\,\ref{#1}}

\def\eq#1{Eq.\,(\ref{#1})}

\def\fig#1{Fig.\,\ref{#1}}

\def\Int#1#2#3{\int_{#1}^{#2}\!{\rm d}#3\,}

\def\C60{C$_{60}$}
\def\au{{\rm au}}\def\ph{{\rm ph}}

\def\final#1{\widetilde{#1}}
\def\ddt#1{\dot#1}

\begin{document}

\title{Prevailing features of X-ray induced molecular electron spectra\\ revealed with fullerenes}
\author{Abraham Camacho Garibay}\author{Ulf Saalmann}\author{Jan M. Rost}
\affiliation{Max Planck Institute for the Physics of Complex Systems\\
             N\"othnitzer Stra{\ss}e 38, 01187 Dresden, Germany }

\begin{abstract}\noindent
Intense X-ray photo-absorption from short and intense pulses by a molecule triggers complicated electron and subsequently ion dynamics leading to photo-electron spectra which are difficult to interpret. Illuminating fullerenes offers a way to separate out the electron dynamics since the cage structure confines spatially the origin of photo and Auger electrons. Together with the sequential nature of the photo processes at intensities available at X-ray free electron lasers, this allows for a remarkably detailed interpretation of the photo-electron spectra as we will demonstrate. The general features derived can serve as a paradigm for less well-defined situations in other large molecules or clusters.
\end{abstract}

\maketitle
\begin{figure}[b!]
\centerline{\includegraphics[width=0.7\columnwidth]{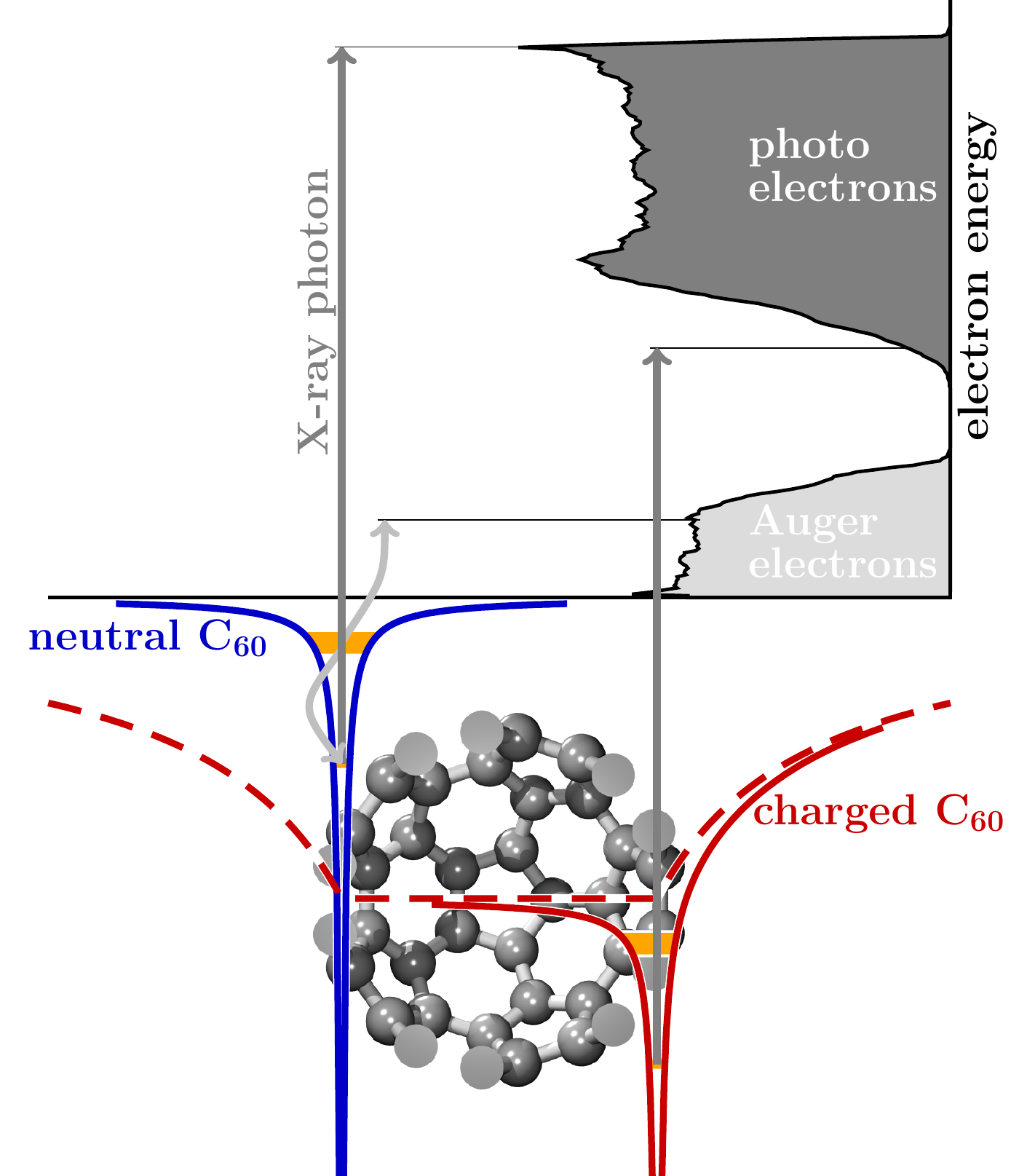}}
\caption{Sketch of the formation of broad electron spectra.
The highest energy is observed for photo-ionization of core electrons from neutral C$_{60}$ (blue solid line on the left).
For a charged C$_{60}$ (red solid line on the right) the energy is reduced due to the background potential (red dashed line).
 }
\label{fig:sketch}
\end{figure}

\noindent
An intense X-ray pulse as available from free-electron lasers \cite{chul+10,bumo+13} triggers multiple electron emission from a large molecule. The resulting electron spectrum is quite complex since it contains photo electrons, Auger electrons and electrons evaporating from an almost thermalized nano-plasma cloud trapped by a positive background charge in the molecule, formed as a consequence of the many photo and Auger electrons leaving the system \cite{saro02,zila+09,both+10,arfe11}.
This charge is considerably larger than in synchrotron experiments \cite{bigi+13}
and along with the corresponding attractive potential depends on time.
The latter builds up quickly with the departure of photo and Auger electrons while it becomes weaker upon expansion induced by ion-ion repulsion. The charging and the expansion of the ionic backbone crucially influences through this potential the electron spectrum since it determines the energy each individual electron carries after leaving the molecule, cf.\ \fig{fig:sketch}.

In this situation it is very desirable to analyze a molecule which exhibits all the features described yet offers a chance for simplification due to its symmetry. An ideal candidate is the \C60 fullerene which has sixty identical carbon atoms on a spherical shell such that each C-atom has the same distance to the molecular center. Moreover, fullerenes can be easily produced in the gas phase as well as in the form of powder samples in the crystalline phase providing an ideal target for experiments with free-electron laser beams. First intriguing femtosecond X-ray diffraction data of crystalline \C60 have been obtained recently \cite{abdi+12}.
%%%%%%%%%%%%%%%%%%%
%%%%%%%%%%%%% 
 
Here, we provide insight into the electron spectrum of \C60 through a theoretical approach which step by step adds complications to the dynamics of the fullerene, ultimately providing a realistic spectrum whose features become then understandable as remnants of the artificially simplified dynamics introduced before. To this end we will first consider one photo electron (and accompanying Auger electron) per atom without trapping and plasma formation and discuss the different shapes of the spectrum obtained from rate equations as a function of the ratio of Auger rate $\gamma$ and photo-ionization rate $\kappa$. The ratio $\gamma/\kappa$ can be varied by changing the intensity of the X-ray pulse. Of course, we keep all ions fixed for this synthetic spectrum.
In a next step we provide the full electron spectrum of \C60 for multi-channel ionization and Auger events determined by a molecular dynamics calculation, but still for fixed ions. In a final and last step we present full dynamical electron spectra including ion motion, starting from very short but realistic pulses (1\,fs duration, $10^{19}$\,W/cm$^2$ intensity) to make contact with the fixed-ion spectra. We will see that ion motion significantly changes the spectra for longer pulses (up to 15\,fs). 
We have refrained from averaging our results over the laser focus for clarity and also,
since in principle simultaneous imaging \cite{goad+12} of the targets may allow in the future to determine accurately under single-shot conditions the intensity with which the molecule was hit.

Dealing with X-rays of more than 1\,keV, we will only consider the photo-ionization of the 1s carbon electrons \cite{yeli85}. We have checked that photo-ionization of electrons from the valence shell in \C60\ is in comparison negligible. They account for less than 5\,\% of the total photo-absorption cross section. 
Auger processes are described on the basis of a molecular Auger lifetime of 6\,fs, a typical value for carbon in molecular environment \cite{caha+00,scsa+04}. 
Photo-absorption as well as Auger-decay processes are modeled with a Monte Carlo procedure. We include all possible transitions (ss, sp, pp) with branching ratios according to the atomic case \cite{ha88} and the instantaneous valence-shell occupations \cite{disa+13}. 
Note that radiative decay (fluorescence) of Carbon core-holes in a molecular environment is about two orders of magnitude slower that non-radiative (Auger) decay \cite{ha88}.

The probability that an electron is emitted with energy $E$ is given by an integral \cite{ro98} over emission times $P(E)=\int\!{\rm d}t\,p(t)\,\delta\big(E{-}E'(t)\big)$ with $p(t)$ the ionization rate at time $t$ and $E'(t)$ the final energy of an electron released at that time.
The energy $E'$ is time-dependent due to the formation of a background potential mentioned before and given explicitly in \eq{eq:eoft}.
Considering various ionization channels $x$, which will be specified below, and writing the ionization rate in terms of $n(t)$, the number of electrons released up to time $t$, the spectrum reads
\begin{subequations}\label{eq:spec}\begin{align}
P(E) & = \sum_{x}\Int{}{}{t} \ddt{n}_{x}(t)\;\delta\big(E{-}E'_{x}(t)\big)
\\
& = \sum_{x}\left|\frac{\ddt{n}_{x}(t)}{\ddt{E}'_{x}(t)}\right|_{t=t'_{x}(E)}
\end{align}\end{subequations}
whereby Eq.\,(\ref{eq:spec}b) follows from properties of the $\delta$-function \cite{ro98} with $t'_{x}(E)$ being the inverse function of $E'_{x}(t)$.
Time derivatives are denoted with a dot.
The final energies of the electrons depend on the charge of all other ions in the \C60 at the instant of their creation.
For the discussion it is convenient to consider an overall charge $Q(t)$ being homogeneously distributed over a sphere with radius $R(t)$.
The corresponding potential is shown in \fig{fig:sketch} as dashed line.
Thereby the final energies are
\begin{subequations}\label{eq:eoft}\begin{align}
E_{x}(t)
& =E_{x}^{\star}+V(t)\\
V(t) & =-\frac{Q(t)}{R(t)}
,\quad
Q(t) = \sum\nolimits_{x}n_{x}(t)\,, \label{eq:eoftb}
\end{align}\end{subequations}
where 
$E_{x}^{\star}$ are the respective excess energies, which would be measured, if an electrons is released from channel $x$ of a single carbon atom.
The excess energies appear as characteristic energies in the spectrum
\begin{subequations}\label{eq:en}\begin{align}
E^\star_\ph & = E_{\omega}-B_{0} \label{eq:enph}
\\
E^\star_\au & = A_{1} \label{eq:enau}
\\
E_\ph^{\rm min} & = E_{\omega}-B_{j_{\rm max}}- \final{Q}/R_{0}\label{eq:enphmin}
\end{align}\end{subequations}
with $B_{j}$ the binding energies of the 1s core in C$^{j+}$
and $A_{j}$ the energies of the valence electrons released during the Auger decay $\alpha_j$ of a core-hole ion C$^{j+}$. 
Indeed there are various Auger decay channels with different energies for a certain charge state $j$. Since those differences are small on the energy scale considered here, we use in the discussion as $A_{j}$ the average of those energies corresponding to $\alpha_j$.
The photo and Auger excess energies (\ref{eq:en}a,b) are ``upper limits'' and correspond to the respective atomic carbon lines.
The third energy \eqref{eq:enphmin} marks a ``lower limit''. It is observed for the last photo electron which is released from C$^{{j_{\rm max}}+}$ and an overall charge $\final{Q}=Q(t{\to}\infty)$.
Here and in the following we denote final variables with a tilde.
As mentioned before, for the moment we neglect ionic motion and set $R(t)\,{=}\,R_{0}\,{=}\,3.52$\,\AA.

%%%%%%%%%%%%%%%%%
\begin{figure}[b]
\begin{center}
\includegraphics[width=0.7\columnwidth]{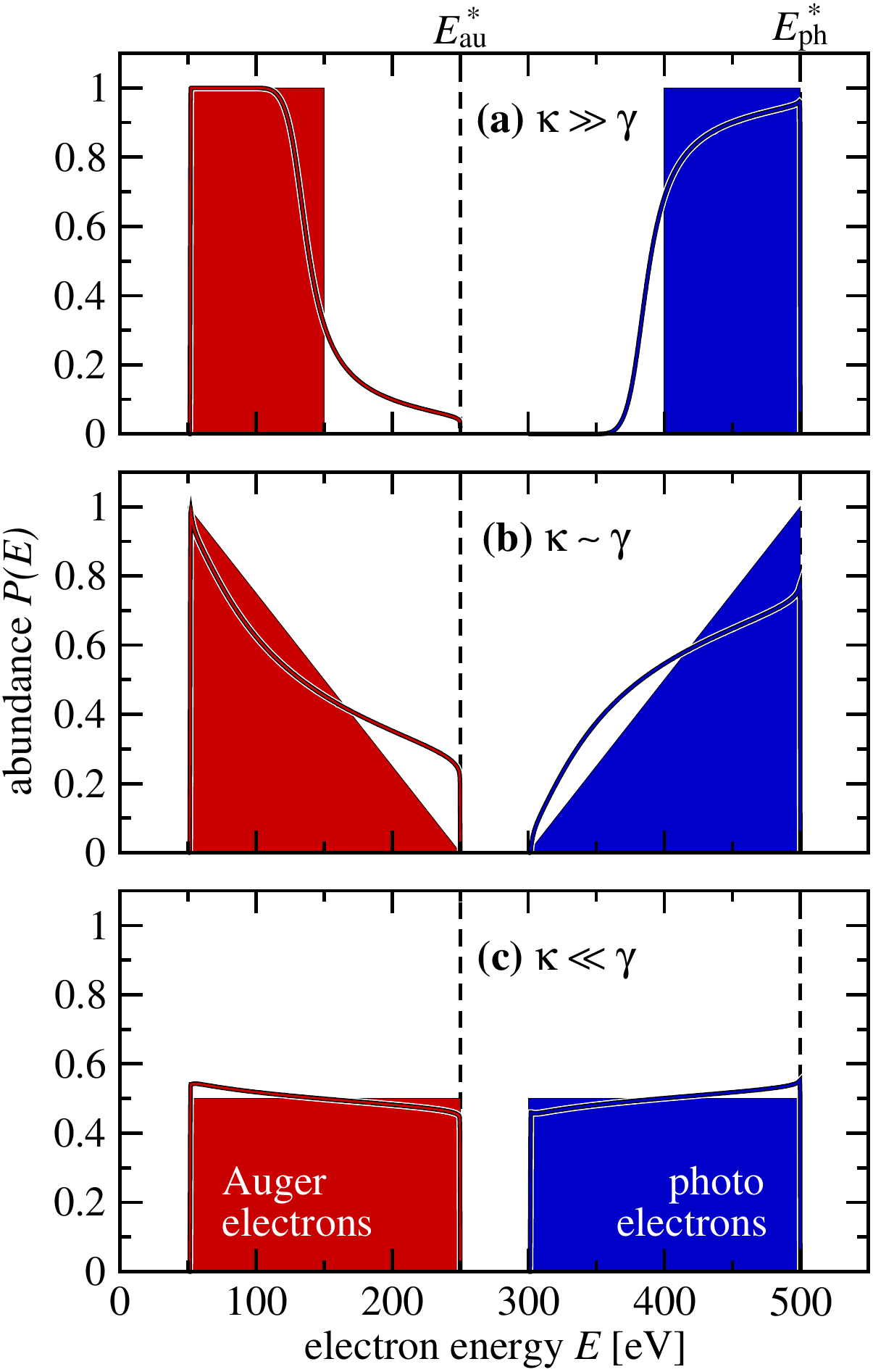}
\caption{Spectra for photo (\textsl{blue}) and Auger (\textsl{red}) electrons for three scenarios (a)--(c) corresponding to different ratios of photo-ionization rate $\kappa$ to Auger rate ($\gamma{=}1$).
From top to bottom $\bar\kappa/\gamma=10,\,1,\,1/10$ with pulse durations $T=2/\bar\kappa=2/10,\,2,\,20$, respectively.
The atomic excess energies of $E^\star_\au{=}250$\,eV and $E^\star_\ph{=}500$\,eV are marked by the dashed lines.
Shaded areas mark the results for rectangular pulses, in (a) and (c) for the limiting case $\kappa{\to}\infty$ and $\gamma{\to}\infty$, respectively, in (b) for the special case of a fixed ratio $\kappa/\gamma = 2$.
}
\label{fig:spec-model}
\end{center}
\end{figure}%
%%%%%%%%%%%%%
We consider in our first simple step only one photo electron per (fixed) atom in C$_{60}$ and the subsequent Auger electron. This leads to the coupled rate equations for the 
time-dependent numbers of neutral atoms $k$, singly- and doubly-charged ions $k^{+}$ and $k^{++}$,
\begin{subequations}\label{eq:model}\begin{align}
\ddt{k}(t) & = -\kappa(t)\;k(t)
\\
\ddt{k}^{+}(t) & = \kappa(t)\;k(t)-\gamma\;k^{+}(t)
\\
\ddt{k}^{++}(t) & = \gamma\;k^{+}(t)\,.
\end{align}\end{subequations}
The initial conditions ($t{\to}{-}\infty$) for neutral \C60 are $k{=}60$ and $k^{+}{=}k^{++}{=}0$. From the ion numbers follow the number of photo-electrons $\ddt{n}_\ph(t)=-\ddt{k}(t)$ and of Auger electrons $\ddt{n}_\au(t)=\ddt{k}^{++}(t)$. 

To simplify things even more we consider first a rectangular pulse with a constant $\kappa$, which allows an explicit solution of the rate equations \eqref{eq:model}. 
The resulting spectrum is shown in \fig{fig:spec-model} for $E_\ph^{\star}=500\,$eV and 
various ratios $\kappa/\gamma$.
We have set $\gamma{=}1$ and chosen $\kappa$ accordingly. 
The pulse duration $T$ was adapted in order to absorb 25 photons, i.\,e., for $t{\to}{+}\infty$ it is $\final{n}_\ph\,{\approx}\,25$ and, of course, $\final{n}_\au\,{=}\,\final{n}_\ph$.

For very fast photo-ionization ($\kappa\,{\to}\,\infty$) all photo electrons produced by the laser pulse have escaped before the first Auger electron is emitted. Consequently, the last photo electron has to escape against a potential of 
$\final{V}\,{\approx}\,{-}100\,$eV, which results from the positive charge of $\final{Q}=\final{n}_\ph$ accumulated through photo-ionization. This explains the rectangular high-energy spectrum in \fig{fig:spec-model}a. 
Such plateau-like spectra have been seen \cite{both+08} and explained \cite{gnsa+11gnsa+12a} before for valence ionization of atomic clusters.
Indeed their appearance is quite general \cite{arfe11,mo09}.
The final energy of the first Auger electron is reduced from $E_\au^{\star}=250$eV to $E_\au^{\star}\,{-}\,\final{V}$, while those of the last one is reduced by $2\final{V} \approx-200$\,eV, since now the total positive charge is $\final{Q}=2\final{n}_\ph$. This explains the low-energy rectangular spectrum for energies $E=50\ldots150$\,eV. 

For the spectrum in \fig{fig:spec-model}c the Auger decays are instant ($\gamma\,{\gg}\,\kappa$). This means that in-between two photo-ionization events an Auger event takes place. Therefore, the background charge for the next photo electron increases by $\delta Q =2$, stretching the interval of the photo-electron spectrum by a factor of 2 to energies $E=300\ldots500$\,eV. The same is true for the Auger spectrum since each photo-electron is accompanied by an Auger electron.
The overall shift of about 250\,eV is due to a lower excess energy $E^{\star}_{\au}$. 

Clearly, if both photo-ionization and Auger decays are decoupled due to very different rates, one expects flat spectra. This changes if both rates are similar. 
In particular for a fixed ratio $\kappa/\gamma = 2$, the electron spectrum takes a special form. In this case the spectra become triangular as shown in \fig{fig:spec-model}b. 
This particular shape can be obtained analytically from \eq{eq:spec}.
The spectra with a time-dependent photo-ionization rate $\kappa(t) = \bar\kappa\sqrt{6/\pi}\exp(-(t/T)^{2})$ following a realistic Gaussian pulse can be understood as deviations from the idealized shapes just discussed for constant photo-absorption in the three limiting cases of ratios $\kappa/\gamma$, cf.\ \figs{fig:spec-model}a--c.
Here we compare the mean photo-ionization rate $\bar\kappa$ \footnote{The mean photo-ionization rate is defined by requiring that the 0th and 2nd moment, $\int{\rm d}t\,\kappa(t)$ and $\int{\rm d}t\,t^{2}\,\kappa(t)$, respectively, agree with the corresponding values for a rectangular pulse with a fixed photo-ionization rate, $\bar\kappa \bar T$ and $\bar\kappa {\bar T}^{3}/12$, respectively.} with the Auger rate $\gamma$.

%%%%%%%%%%%%%%%%%%%
\begin{figure}[b]
\begin{center}
\includegraphics[width=0.9\columnwidth]{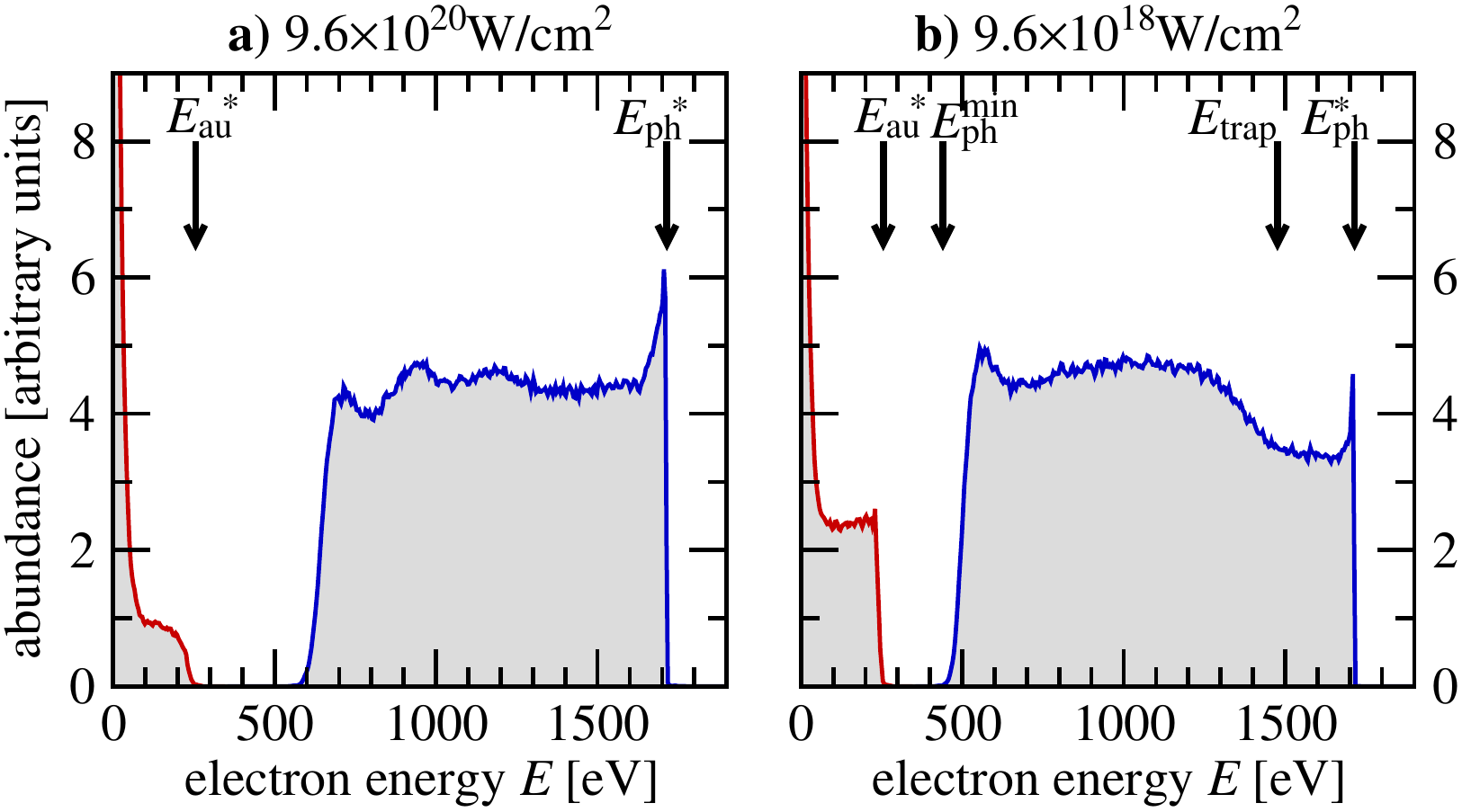}
\caption{Electron spectra for fixed ions generated with an X-ray pulse of 2\,keV photons. 
The peak intensities are $9.6\times 10^{20}$W/cm$^2$ and $9.6\times 10^{18}$W/cm$^2$ for scenarios (a) and (b), respectively. 
From the area under the photo electron spectrum in the interval $[E_\ph^{\rm min},E^\star_\ph]$ follows the number of photo electrons: $n_\ph^\mrm{(b)} = 238, n_\ph^\mrm{(a)} = 217$.}
\label{fig:scenarios}
\end{center}
\end{figure}%
%%%%%%%%%%%%%

As a next step of complication we discuss the full electron spectrum for fixed ions resulting from many-channel electron dynamics including double core-hole and plasma formation for
the scenarios (a) and (b) from \fig{fig:spec-model}
with Gaussian pulses delivering photons of $E_{\omega}{=}\,2$\,keV energy.
Results have been obtained with a molecular dynamics propagation of released electrons and the
parameters are chosen such that more than 200 photo electrons are produced to activate all subsequent channels down to photo-ionization of the ground state of C$^{5+}$ with $B_5 = 476\,$eV entering \eq{eq:enphmin}. A quick glance on \fig{fig:scenarios} reveals that $E^\star_\au$ and $E^\star_\ph$ are the same for both scenarios as expected.

In scenario (b) the photo-ionization rate is small enough that on average an Auger electron is produced after each photo electron. A characteristic step appears at energy $E_{\rm trap} = E_{\omega}\,{-}\,B_{0}\,{-}\,A_{1} = 1476$\,eV (see \fig{fig:scenarios}b) separating 
the high-energy part from the low-energy part in the photo-electron spectrum. In the former after each photo electron an Auger electron escapes increasing the background charge by $\delta Q{=}2$ while in the latter the Auger electron is trapped such that $\delta Q{=}1$. Trapping occurs if the background potential is deep enough such that
 $Q_{\rm trap}/R = A_1 = 239\,$eV, the kinetic energy of the Auger electron produced in the Auger decay of C$^+$ with a 1s hole. Consequently, $Q_{\rm trap}/2 \approx 29$ is the number of escaped Auger electrons, provided this step at $E_{\rm trap}$ exists in the spectrum. 
With the 29 escaped Auger electrons and the 238 photo electrons
 we estimate from \eq{eq:enphmin} $E_\ph^{\rm min} = 440\,$eV in good agreement with \fig{fig:scenarios}b.

%%%%%%%%%%%%%%%%%%
\begin{figure}[b]
\begin{center}
\includegraphics[scale=0.55]{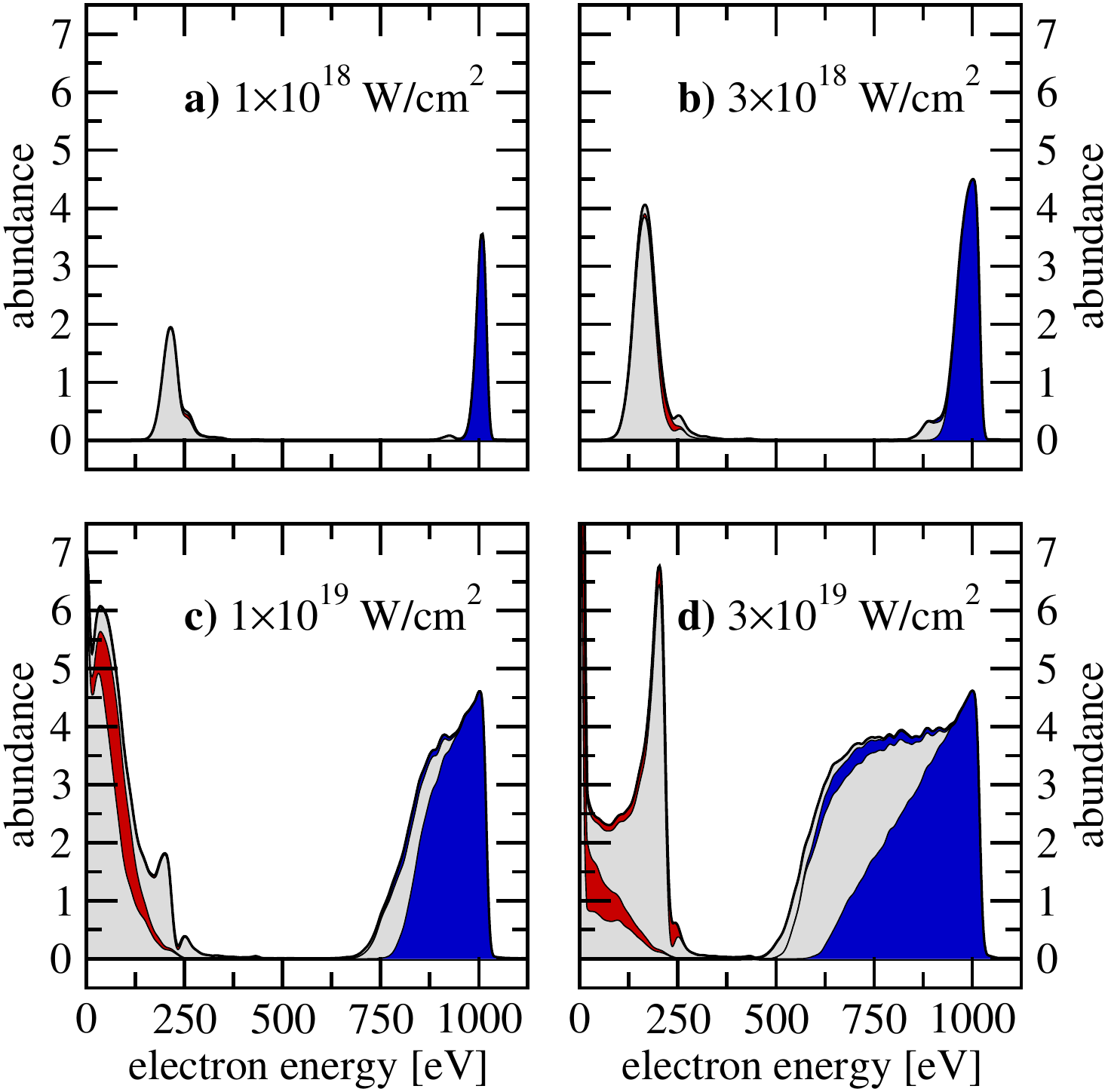}
\caption{(a-d) Electron energy spectra from C$_{60}$ exposed to a short pulse ($T=1$\,fs) of $E_{\omega} = 1.3$\,keV photons for various intensities. Photo-electron contributions from $C^{2j}$ are indicated in blue,
from corresponding Auger electrons in red, and from all odd charged ions $C^{2j+1}$ in grey. The area under the photo-electron spectra provides the number of photo electrons: $\big\{n_\ph^\mrm{(a)},n_\ph^\mrm{(b)},n_\ph^\mrm{(c)},n_\ph^\mrm{(d)}\big\} = \{6,17,48,98\}$.}
\label{fig:spec-intens}
\end{center}
\end{figure}%
%%%%%%%%%%%%%
The spectrum of scenario (a) with $\kappa\,{\gg}\,\gamma$ lacks $E_{\rm trap}$, since fast photo-ionization builds up a large $Q$ before Auger electrons are produced. 
Dominant photo-ionization early in the pulse is confirmed by the triangular shape of the photo electron spectrum at its blue edge ($E\,{\lesssim}\,E_{\ph}^{\star}$). It forms if exactly two channels contribute whose
rate has the ratio 2. This is the case for photo-ionization of C and C$^+$ with a single core-hole without significant influence of other channels, e.g., Auger escape. Given that situation, we can read-off from \fig{fig:scenarios}a that $E_\ph^{\rm min}\approx 600\,$eV and estimate with the help of \eq{eq:enphmin} and the 217 photo electrons that about 10 Auger electrons have escaped.

%%%%%%%%%%%%%%%%%%
\begin{figure}[t]
\begin{center}
\includegraphics[scale=0.55]{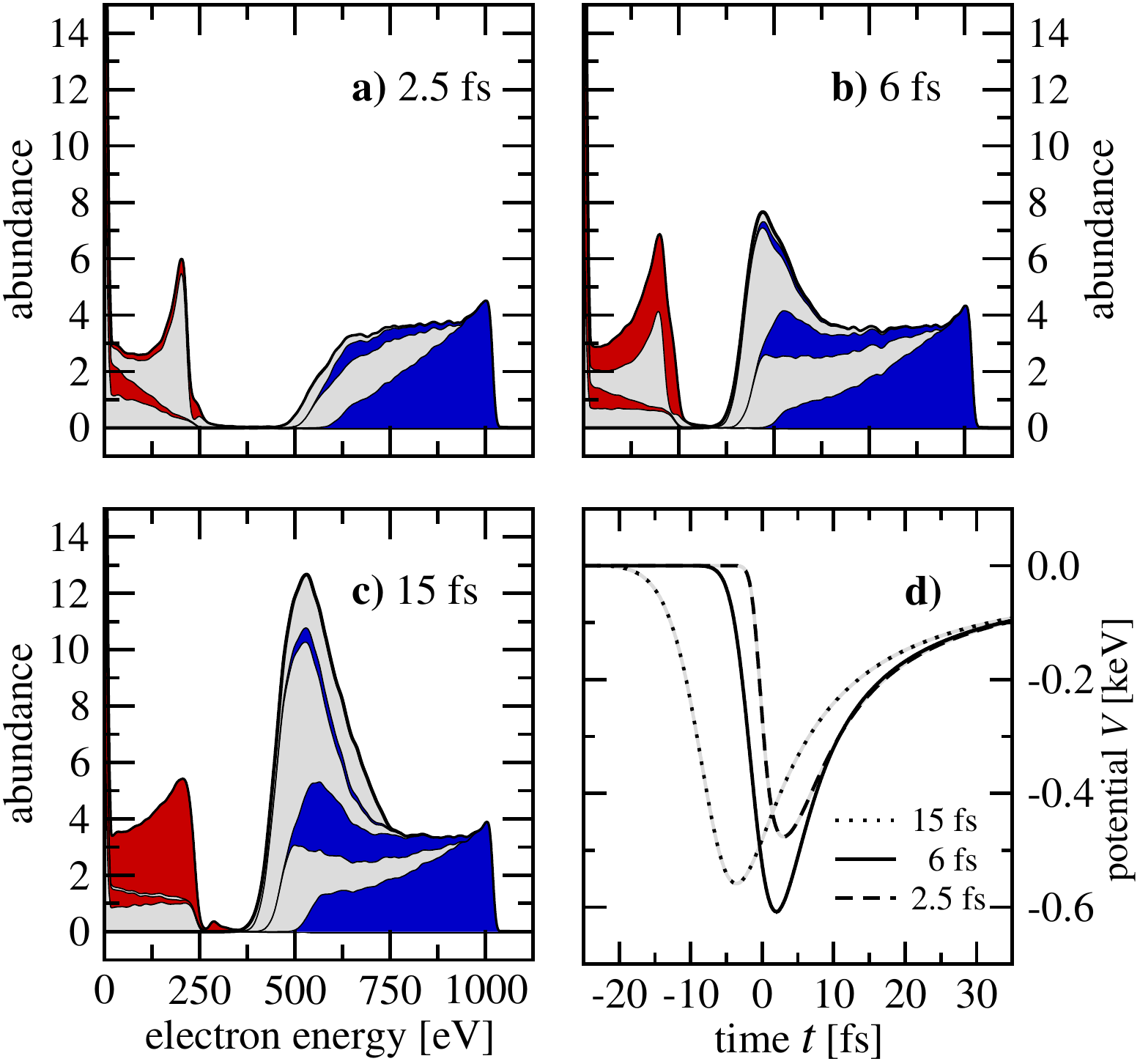}
\caption{(a--c) Electron energy spectra as in \fig{fig:spec-intens} but for a fixed peak 
intensity of $I=10^{19}$W/cm$^2$ and various pulse lengths $T$. 
Note that the spectrum for $T{=}1$\,fs with the same intensity $I$ is shown in \fig{fig:spec-intens}c. 
Panel (d) shows the time dependent background potential $V(t)$, given by \eq{eq:eoftb}, for the same pulse parameters.
The number of photo electrons are $\big\{n_\ph^\mrm{(a)},n_\ph^\mrm{(b)},n_\ph^\mrm{(c)},n_\ph^\mrm{(d)}\big\} = \{48,96,155,208\}$.
}
\label{fig:spec-pulses}
\end{center}
\end{figure}%
%%%%%%%%%%%%% 
We are now in a position to understand the truly complex electron spectra resulting from full electron and ion dynamics when illuminated by an intense X-ray pulse of $E_\omega = 1.3$\,keV photons.
The highest intensities used are realistic, assuming $10^{13}$\,photons per pulse focused onto 1\,$\mu$m$^{2}$ which gives roughly $I_{\rm max} \approx (2/{T_{\rm fs}){\times}10^{20}}\mbox{W}/\mbox{cm}^{2}$. 
To make contact with the frozen ions discussed so far we consider first in \fig{fig:spec-intens} a $T\,{=}\,1$\,fs pulse. 
For weak intensity, i.\,e., only a few electrons ionized, the spectrum in \fig{fig:spec-intens}a shows peaks slightly red-shifted to the atomic photo-electron and Auger-electron energies $E^\star_\ph$ and $E^\star_\au$. 
For further increasing intensity the familiar \emph{plateau\/} for the photo electrons develop due to the generation of background charge $Q$ and for the same reason the Auger peak gets an increasing red-shift. 
As in \fig{fig:scenarios} the energies $E_\ph^{\rm min}$ are consistent with the number of photo and Auger electrons which have escaped and the relevant excess energies $B_j$, cf.\ the figure caption. 
However, the Auger electrons reveal a new feature, namely the re-appearance of a peak close to the atomic line $A_3=224\,$eV for high intensity, see \figs{fig:spec-intens}c,d. 
The reason for this behavior is that the background potential \eq{eq:eoftb} is weak at early \emph{and\/} at late times when the fullerene is either weakly charged ($Q{\approx}0$) or is exploded ($R{\to}\infty$), see \fig{fig:spec-pulses}d. 
Indeed, further analysis reveals that the $
\alpha_3$-related peaks in \figs{fig:spec-intens}c,d require a preceding Auger decay $\alpha_2$ due to the short pulse (1\,fs) which considerably delays the $\alpha_3$-decay. 

The non-monotonic time dependence of the background potential together with the
different emission times of the electrons can change the shape of the electron spectra qualitatively.
This is further corroborated with \fig{fig:spec-pulses}a--c,
where one sees drastic changes also in the photo-electron spectrum for longer pulses. The dominant effect is that electrons emitted late in or even after the laser pulse loose less and less energy turning the formation of a plateau into a pile up of electrons in a small energy interval, particularly striking for the (late) photo electrons from C$^{j+}$ with $j{>}2$ in \fig{fig:spec-pulses}c. 
 
In summary we have analyzed the electron spectra of fullerenes following irradiation of short and intense X-ray pulses.
These spectra reveal general features which will also allow one to interpret spectra of less symmetric large molecules.
In particular triangular shapes will appear whenever two ionization channels are dominantly involved where one is twice as likely as the other, as it is the case with the photo-ionization of a filled (1s$^2$) and half-filled (1s) core shell.
Moreover, we have established the onset $E_\ph^{\rm min}$ of the photo electron spectrum as characteristic energy which allows one to determine directly the number of escaped Auger electrons in the case of \C60. 
Finally, peaks in the Auger spectra have to be interpreted carefully since they may not point to a unique decay process but to Auger electrons emitted from the molecule at rather different times with the same energy due to the non-monotonic change of the background potential.
These features should prevail even if photon energies and Auger energies lead to overlapping photo and Auger electron spectra as well as post collision interaction, additional complications which will be investigated in future work.

This work was supported by the COST Action XLIC (CM\,1204) and the Marie Curie Initial Training Network CORINF.

\end{document}